\documentclass{iau} 
\usepackage{graphicx}

\title[Shocks and UV radiation around protostars] 
{Shocks and UV radiation around low-mass protostars: the Herschel-PACS legacy}

\author[Agata Karska, Michael J. Kaufman, Lars E. Kristensen et al.]   
{Agata Karska$^{1,2,3}$,  Michael J. Kaufman$^4$, Lars E. Kristensen$^5$, Ewine F. van Dishoeck$^{3,2}$
 \and WISH, DIGIT, and WILL teams}

\affiliation{$^1$Centre for Astronomy, Faculty of Physics, Astronomy and Informatics,
 Nicolaus Copernicus\\ University,
Grudziadzka 5, 87-100 Torun, Poland \\ email: {\tt agata.karska@umk.pl} \\[\affilskip]
$^2$Max-Planck Institut f\"{u}r Extraterrestrische Physik (MPE), \\ 
Giessenbachstr. 1, D-85748 Garching, Germany \\[\affilskip]
$^3$Leiden Observatory, Leiden University, \\ 
P.O. Box 9513, 2300 RA Leiden, The Netherlands \\[\affilskip]
$^4$Department of Physics and Astronomy, San Jose State University, \\ 
One Washington Square, San Jose, CA 95192-0106, USA \\[\affilskip]
$^5$Centre for Star and Planet Formation, Niels Bohr Institute, University of Copenhagen, \\ 
{\O}ster Voldgade 5-7, DK-1350 Copenhagen K, Denmark} 

\pubyear{2017}
\volume{332}  
\jname{Astrochemistry VII - Through the Cosmos from Galaxies to Planets}
\editors{M.~Cunningham, T.~Millar \& Y. Aikawa, eds.}
\begin{document}
\maketitle

\begin{abstract}
Far-infrared spectroscopy reveals gas cooling and its underlying heating due to physical 
processes taking place in the surroundings of protostars. These processes are reflected in both the 
chemistry and excitation of abundant molecular species. Here, we present the \textit{Herschel}-PACS far-IR 
spectroscopy of 90 embedded low-mass
  protostars from the WISH \cite[(van Dishoeck et al. 2011)]{vD11},
 DIGIT \cite[(Green et al. 2013)]{Gr13}, and WILL surveys \cite[(Mottram et al. 2017)]{Mo17}.
  The $5\times5$ spectra covering the $\sim50''\times50''$ field-of-view include rotational
transitions of CO, H$_2$O, and OH lines, as well as fine-structure [O I] and [C II]
in the $\sim$50-200 $\mu$m range. The CO rotational temperatures (for $J_\mathrm{u}\geq14)$ 
are typically $\sim$300 K, 
with some sources showing additional components with temperatures as high as $\sim$1000 K.
The H$_2$O / CO and H$_2$O / OH flux ratios are low compared to stationary shock models, 
suggesting that UV photons may dissociate some H$_2$O and decrease its abundance.
Comparison to C shock models illuminated by UV photons shows good agreement between 
the line emission and the models for pre-shock densities of $10^5$ cm$^{-3}$ and UV fields 0.1-10 times
the interstellar value. The far-infrared molecular and atomic lines are the unique
diagnostic of shocks and UV fields in deeply-embedded sources.
\end{abstract}

\firstsection 
\section{Introduction}
During the early formation of low-mass protostars, the mass accretion rates are high and 
thus, the feedback from the protostar on its surrounding is most spectacular. The launching of  
jets and winds creates outflow cavities and generates shock waves 
propagating in the envelope. Shocks compress and heat the gas but may also produce UV 
photons, which penetrate to large distances due to low densities and scattering in the 
outflow cavities. Characterising these processes requires spectroscopy in the far-infrared (IR), 
where cooling of the warm, dense gas is very efficient \cite[(Kaufman \& Neufeld 1996)]{KN96}.
The question remains, how do these feedback processes influence the initial conditions (both physics and chemistry) 
of star and planet formation.

In this paper, we briefly summarize the survey of 90 deeply-embedded low-mass 
protostars observed as part of the three large \textit{Herschel} programs: `Water in Star forming regions with Herschel' 
\cite[(WISH, van Dishoeck et al. 2011)]{vD11}, `Dust, Ice, and Gas in Time' \cite[(DIGIT, Green et al. 2013)]{Gr13},
 and \lq William Herschel Line Legacy' \cite[(WILL, Mottram et al. 2017)]{Mo17}. All
observations were obtained using the Photodetector Array Camera and Spectrometer 
\cite[(PACS, Poglitsch et al. 2010)]{Po10} that provided spectral maps consisting
of 25 spatial pixels of $\sim9.4''\times9.4''$, corresponding to pixel sizes of $\sim2000\times2000$ AU
 for typical distances of 200 pc. 

\section{Results}
\begin{figure*}[!tb]
  \begin{minipage}[t]{.45\textwidth}
  \begin{center}
       \includegraphics[angle=90,height=6cm]{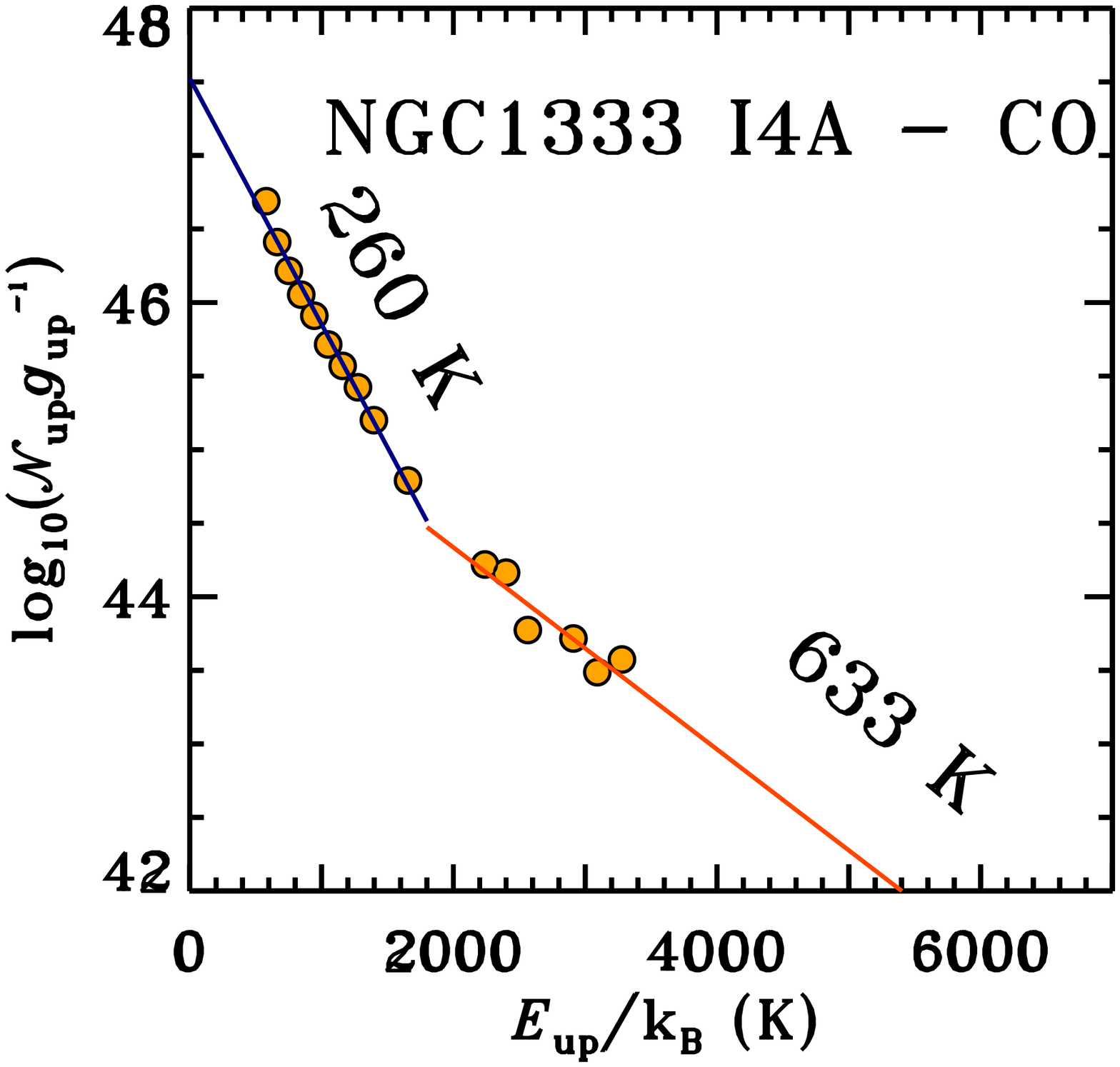} 
           \hspace{-2pt}
   \end{center}
  \end{minipage}
  \hfill
  \begin{minipage}[t]{.45\textwidth}
      \begin{center}
   	   \includegraphics[angle=90,height=6cm]{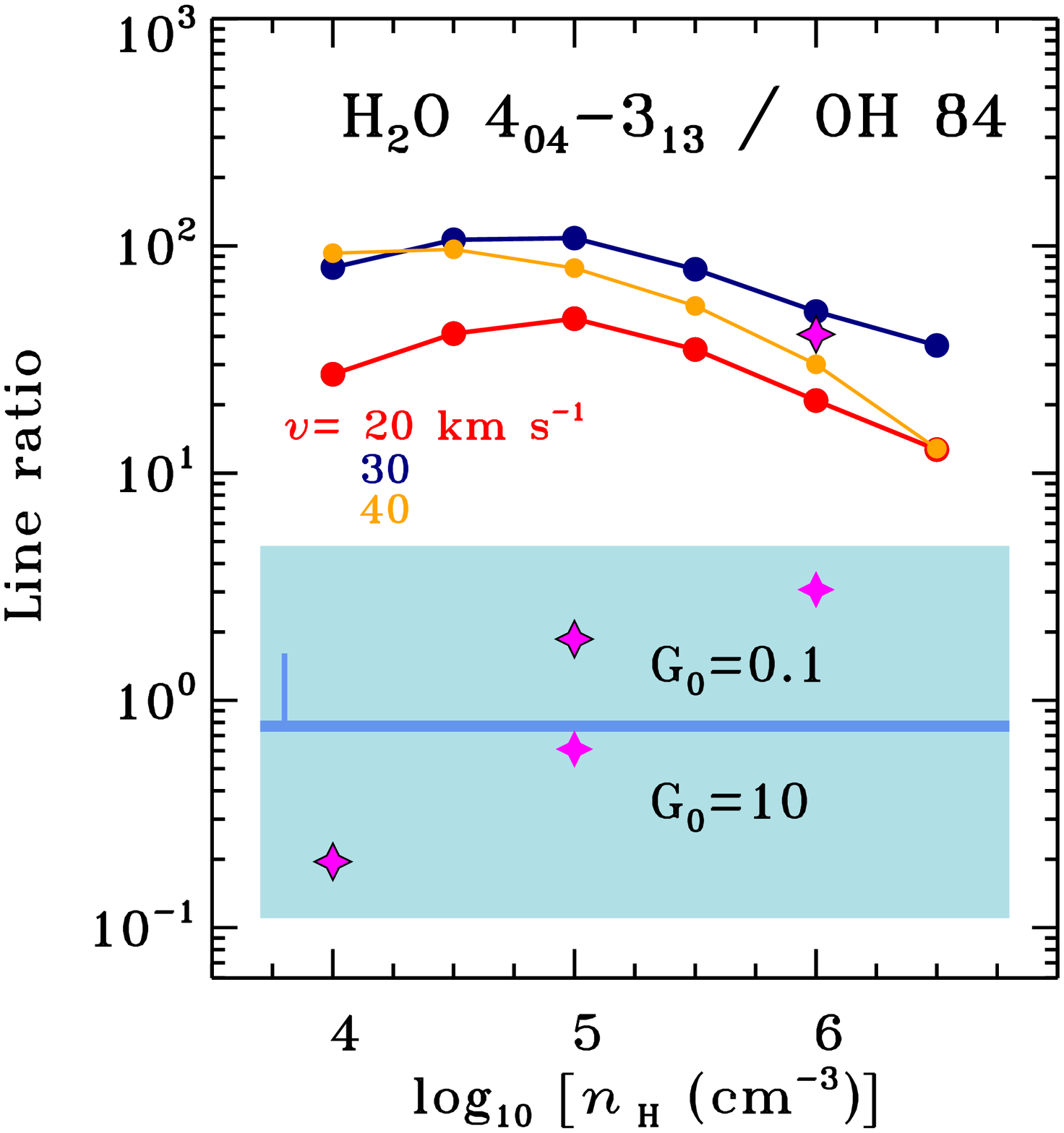} 
   	       \hspace{-2pt} 
      \end{center}
  \end{minipage}
    \hfill
\caption{\label{fig1} \textit{Left:} CO rotational diagram of a Class 0 protostar, NGC 1333 IRAS 4A. 
\textit{Right:} Line ratio of H$_2$O $4_{04}-3_{13}$ at 125 $\mu$m and OH at 84 $\mu$m as a function 
of logarithm of density of the pre-shock gas. The range observed in $\sim$90 protostars is shown as the filled, light-blue 
box, and the models of C shocks are shown as solid lines \cite[(models of shielded shocks, Kaufman 
\& Neufeld 1996)]{KN96} and violet stars \cite[(models of UV illuminated shocks, Melnick 
\& Kaufman 2015)]{MK15}.}
\end{figure*}
Rich molecular line emission is commonly detected in Class 0/I protostars including highly-excited 
H$_2$O ($E_\mathrm{up}\sim 1000$ K) and CO lines ($E_\mathrm{up}\sim 5000$ K), see 
\cite[Herczeg et al. 2012 and Goicoechea et al. 2012]{He12,Go12}. The line fluxes of 
molecular species correlate strongly with each other and show similar extent of emission,
typically following the outflow direction. Multiple 
transitions of CO lines are used to determine rotational temperatures of the gas 
(see left panel of Figure 1). A $\sim300$ K component is universally seen in almost all protostars, 
whereas the hotter gas component shows a broad distribution with a median at 720 K (detected in 27\% of sources).
These two physical components are identified in the spectrally-resolved CO 16-15 lines obtained with HIFI
\cite[(Kristensen et al. 2017)]{K17}. The $\sim300$ K component in the PACS CO ladder corresponds to the 
broad, Gaussian component in the line profiles originating in outflow cavity shocks or disk winds. 
The hotter component is associated with the \lq\lq offset'' component (i.e. offset in velocity) likely produced in the irradiated 
shocks (see Kristensen et al. 2017, for discussion).

The right panel of Figure 1 shows a comparison of the measured H$_2$O / OH line ratios 
and the predictions from stationary (1D plane-parallel) C shock models from \cite[Kaufman \& Neufeld (1996)]{KN96}. 
Clearly, these models overestimate the observed line ratios by 1-2 orders of magnitude.
Low H$_2$O / OH line ratios require partial photodissociation of H$_2$O into 
OH and O. Predictions from C shock models illuminated by UV photons \cite[(Melnick \& Kaufman 2015)]{MK15} 
reproduce the observed ranges of H$_2$O / OH for pre-shock densities of $10^5$ cm$^{-3}$ 
and UV fields 0.1-10 times the interstellar value. These preliminary comparisons show that 
UV radiation influences the chemistry in the inner envelopes around deeply-embedded protostars, as also evidenced by 
observations of light ionised hydrides \cite[(Kristensen et al. 2013, Benz et al. 2016)]{Kr13,Be16}. 
\begin{figure*}[!tb]
  \begin{minipage}[t]{.45\textwidth}
  \begin{center}
       \includegraphics[angle=90,height=4.5cm]{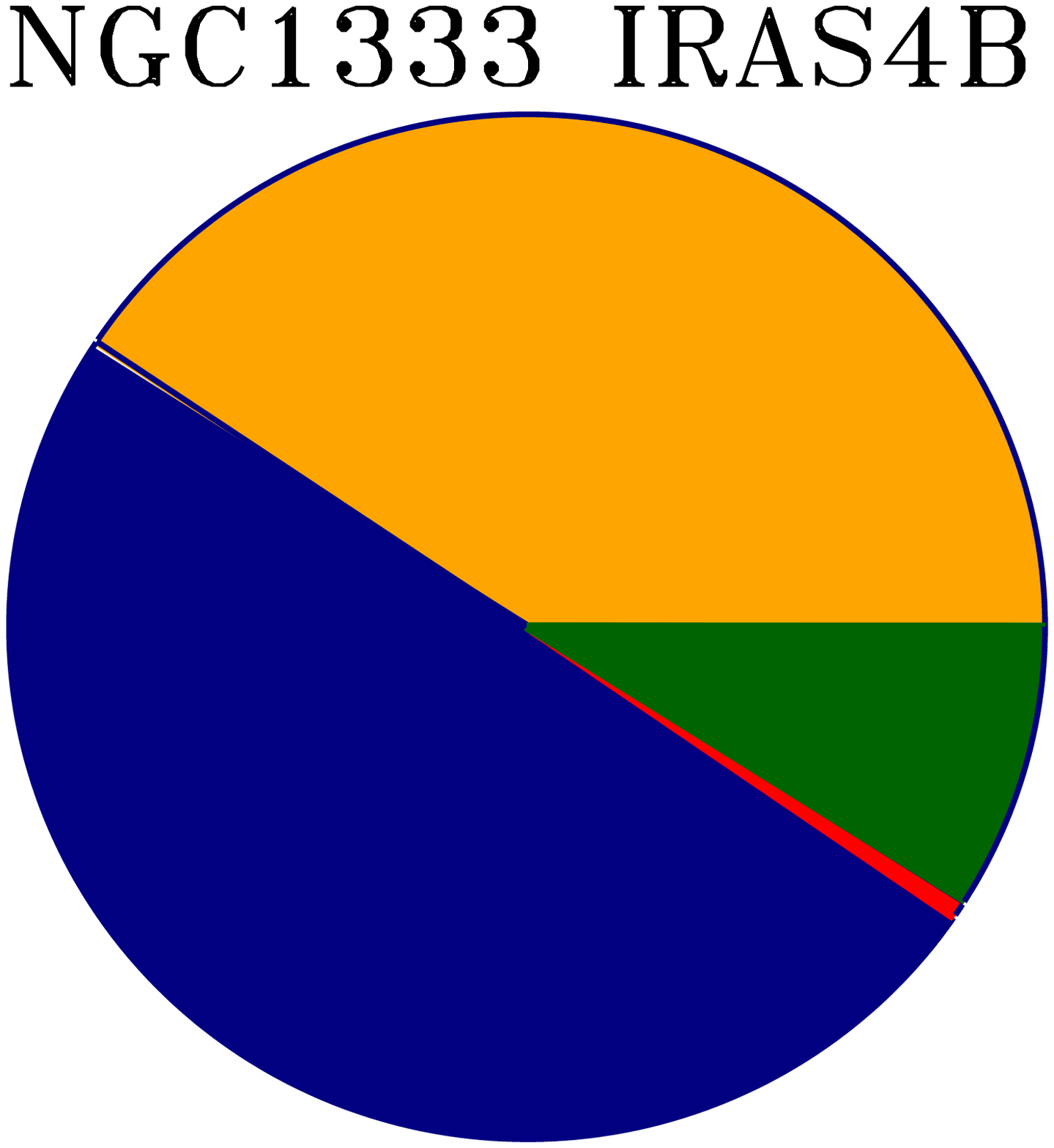} 
           \hspace{-1pt}
   \end{center}
  \end{minipage}
  \hfill
  \begin{minipage}[t]{.45\textwidth}
      \begin{center}
   	   \includegraphics[angle=90,height=4.5cm]{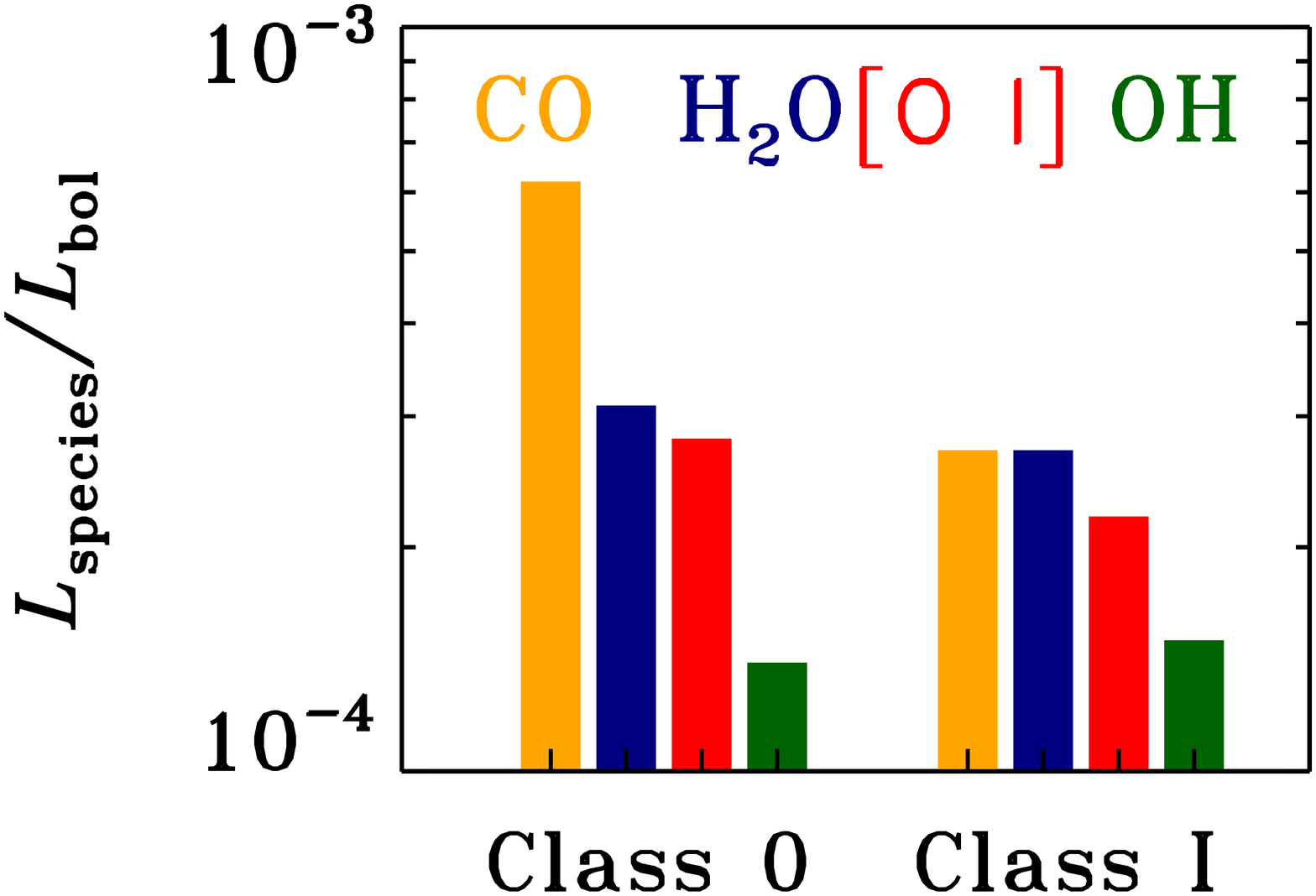} 
   	       \hspace{-1pt} 
      \end{center}
  \end{minipage}
    \hfill
\caption{\label{fig2} \textit{Left:} Fractions of gas cooling contributed by CO (orange), H$_2$O (blue),
         OH (green), and [O I] (red), to the total far-infrared line cooling in the Class 0 protostar NGC 1333 IRAS 4B.
\textit{Right:} Median line cooling in CO, H$_2$O, [O I], and OH 
over bolometric luminosities for Class 0 and Class I sources.}
\end{figure*}

Low abundances of H$_2$O are also confirmed by relatively low total line cooling in H$_2$O 
versus other molecules at far-IR wavelengths. \cite[Kaufman \& Neufeld (1996)]{KN96} predict that 
H$_2$O should account for 70-90\% of total molecular line cooling in the far-IR 
\cite[(H$_2$O+CO+OH, Karska et al. 2014)]{Ka14}. At the same time, the observed H$_2$O emission 
is typically below 30\% and at most about 50\% in the most H$_2$O-rich 
source NGC1333 IRAS4B (left panel of Figure 2, data from Herczeg et al. 2012). Clearly, some H$_2$O is missing and the 
UV photodissociation is a possible mechanism to decrease its abundance. 

Far-IR molecular line cooling decreases from Class 0 to Class I (right panel of Figure 2), 
but the differences are significant only in CO and not in H$_2$O (Karska et al. subm.). Proper interpretation 
of these trends requires the analysis of velocity-resolved profiles, which reveal multiple 
kinematic components with varying contributions depending on the transition 
\cite[(e.g., Kristensen et al. 2017)]{K17}. 

The far-IR data presented here demonstrate 
the necessity for excellent space-based facilities to quantify the chemistry and excitation
 of warm molecular and atomic gas. The next major step will occur when the \textit{James Webb Space Telescope}
  (JWST) is launched in 2018; this telescope will provide crucial data on not only the 
  mid-IR cooling species, but most importantly on the dominant gas coolant: 
  molecular hydrogen, H$_2$. Direct measurements of the atomic and molecular
   abundances will be performed and tested against sophisticated shock models.

{\underline{\it Acknowledgement}}. The authors are grateful to G. Herczeg, J. Mottram, 
L. Tychoniec, J. Lindberg, N. Evans II, J. Green, Y.-L. Yang, A. Gusdorf, and N. Siodmiak
for collaborations on this project. This work is supported by the Polish National Science 
Center grant 2013/11/N/ST9/00400.

\end{document}